\documentclass[smallextended]{svjour3}       
\smartqed  
\usepackage{graphicx}

\usepackage{dcolumn}

\usepackage{amsmath,amsthm,amssymb}
\usepackage{subfigure}
\usepackage{floatrow}
\usepackage{mathrsfs}
\def\ep{\varepsilon}

\newtheorem{conj}{Conjecture}
\RequirePackage{fix-cm}
\begin{document}

\title{Is AdS stable?
}


\author{Piotr Bizo\'n
}


\institute{Piotr Bizo\'n \at
              Institute of Physics, Jagiellonian
University, Krak\'ow, Poland\\
Max Planck Institute for Gravitational Physics (Albert Einstein Institute),
Golm, Germany
              \email{piotr.bizon@aei.mpg.de}
}

\date{Received: date / Accepted: date}

\maketitle

\begin{abstract}
It has recently been conjectured that the Anti-de Sitter space is unstable under arbitrarily small perturbations. This article (based on my plenary talk of the same title at the conference GR20 in Warsaw) briefly reviews numerical and analytical evidence supporting this conjecture, putting emphasis on weak turbulence as a driving mechanism of instability.
\end{abstract}
\section*{Introduction}
 Over the past 15 years asymptotically Anti-de Sitter (AdS) spacetimes
   have come to play a central role in theoretical
 physics, primarily
 due to the AdS/CFT correspondence which
 is the conjectured equivalence between string theory on an asymptotically AdS spacetime and a conformally invariant  quantum field theory (CFT)  living on  the boundary of this spacetime \cite{mal}.
                 For strongly coupled CFT, the string dual effectively reduces to classical AdS gravity which makes the following strategy attractive: (i) construct a gravity side of duality,
                        (ii) use an AdS/CFT dictionary to translate the result to the CFT side, and (iii)
                        compare the result to the real-world physics that your CFT is supposed to model. The problem with
this holographic approach to  modeling non-equilibrium processes (like heavy ion collisions) is that our current understanding of the gravity side is very limited in the non-stationary regime. This provides strong motivation for studying the dynamics of asymptotically AdS spacetimes, especially that, regardless of potential AdS/CFT applications,  this is an interesting  problem in classical general relativity on its own.

\section*{Anti-de Sitter space} Anti-de Sitter space
 is the unique
   maximally symmetric Lorentzian manifold with constant negative scalar curvature. In $d+1$ dimensions it can be represented geometrically as the hyperboloid of radius $\ell$
  \begin{equation}\label{quadric}
    X_1^2+\dots+X_{d}^2-U^2-V^2=-\ell^2
  \end{equation}
embedded in the flat $d+2$ dimensional space with  metric
\begin{equation}\label{emb}
ds^2=dX_1^2+\dots+dX_{d}^2-dU^2-dV^2\,.
\end{equation}
 In terms of the parametrization $X=r \omega$ (where $\omega \in S^{d-1}$), $U=\sqrt{r^2+\ell^2}\,\sin(\tau/\ell)$, and  $V=\sqrt{r^2+\ell^2}\,\cos(\tau/\ell)$, the induced metric on the hyperboloid \eqref{quadric} is
\begin{equation}\label{ads-metric}
 g=-(1+r^2/\ell^2)\, d\tau^2 + \frac{dr^2}{1+r^2/\ell^2} + r^2 d\omega^2\,,
\end{equation}
where $d\omega^2$ is the round metric on the  unit $(d-1)$ -- dimensional sphere.
This metric is the solution of  vacuum Einstein's equations
  $G_{\alpha\beta}+\Lambda g_{\alpha\beta}=0$ with negative cosmological constant $\Lambda=-\dfrac{2}{d(d-1)\ell^2}$. The advantage of representing the AdS space by embedding is that its  symmetry group $O(2,d-1)$ is manifest. The disadvantage is that
   the hyperboloid \eqref{quadric} has the topology $S^1\times \mathbb{R}^d$ and the circles $S^1$ are closed timelike lines.
  A simple remedy to this causality violation is to unroll the circle $S^1$ to its covering space $\mathbb{R}$ and thereby pass to the universal covering space of AdS with the topology of $\mathbb{R}^{d+1}$. Henceforth, by the AdS space we shall always mean this universal covering space.

 The AdS space has  peculiar causal properties. To see them, it is convenient to introduce
  dimensionless coordinates $t=\tau/\ell$ and $x=\arctan(r/\ell)$ with range $(t,x)\in \mathbb{R}\times [0,\pi/2)$, in which the metric \eqref{ads-metric} takes the form\footnote{In the following we choose the AdS radius $\ell$ as the unit of length which is equivalent to setting $\ell=1$.}
\begin{equation}\label{ads}
g = \frac{\ell^2}{\cos^2{\!x}}\left(-dt^2 + dx^2 + \sin^2{\!x}\, d\omega^2\right)\,,
\end{equation}
 showing that the AdS space is conformal to half of the Einstein static universe. The conformal infinity $\mathcal{I}=\{x=\pi/2\}$  is the timelike cylinder $\mathbb{R}\times S^{d-1}$ with the boundary metric $g_{\mathcal{I}}=-dt^2+d\omega^2$ (the conformal diagram of AdS is shown in Fig.~1). Even though the spatial distance from any point in the region $0\leq x<\pi/2$ (the `bulk') to the boundary $ \mathcal{I}$ is infinite,  null geodesics get there in finite time  (but infinite affine time so they are future complete). As a consequence of  the timelike spatial (and null) infinity, the AdS space is not globally hyperbolic, that is there is no Cauchy hypersurface. In order to determine of evolution of fields on AdS one has to prescribe -- in addition to initial data on the $t=0$ hypersurface --  suitable boundary conditions at $\mathcal{I}$.
\begin{figure}[h]
\floatbox[{\capbeside\thisfloatsetup{capbesideposition={right,center},capbesidewidth=0.5\textwidth}}]{figure}[\FBwidth]
{\caption{The conformal diagram of AdS space (all angular dimensions have been suppressed). The diagram is endless in the future and past directions.  The light ray sent outwards from the point $(0,0)$ follows the null geodesic $t=x$ and reaches infinity at the point $(\pi/2,\pi/2)$. Beyond this point the evolution of the light ray (as governed by  Maxwell's equations) depends on the choice of a boundary condition at $x=\pi/2$, which is indicated by the question mark.}\label{fig:test}}
{\includegraphics[width=0.3\textwidth]{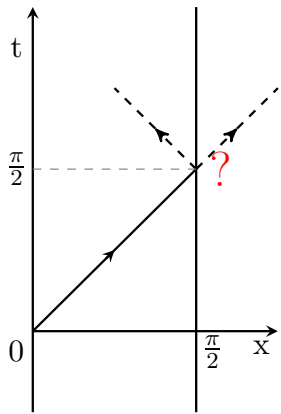}}
\end{figure}

  Spacetimes that  approach the AdS space  at infinity fast enough and have  the same  conformal boundary are called asymptotically AdS spacetimes\footnote{For  precise definitions of asymptotically AdS spacetimes  see, e.g., \cite{am,ht,f,cs}.}. Asymptotically AdS spacetimes may be  very different from the pure AdS in the bulk, in particular may contain horizons. A prototype asymptotically AdS spacetime, besides the AdS space itself, is the AdS-Schwarzschild black hole with the metric
  \begin{equation}\label{schw}
g_S = \frac{1}{\cos^2{\!x}}\left(-A \,dt^2 + A^{-1} dx^2 + \sin^2{\!x}\, d\omega^2\right),\quad A=1-\frac{M (\cos{x})^{d}}{(\sin{x})^{d-2}}\,,
\end{equation}
   where $M>0$ is the mass.  By the positive energy theorem (for globally regular solutions of Einstein's equations with matter satisfying the dominant energy condition),  the AdS space is a ground state among asymptotically AdS spacetimes  \cite{ghhp}, much as Minkowski space is a ground state among asymptotically flat spacetimes \cite{sy,ew}.

    For any ground state the fundamental question is whether it is stable, i.e. do small perturbations of it at $t=0$ remain small for all future times (where `small' is defined in terms of an appropriate norm)? For Minkowski space this question has been answered in affirmative by Christodoulou and Klainerman \cite{ck}, who proved that sufficiently small perturbations not only remain small but decay to zero with time in any compact region (this stronger type of stability is called asymptotic stability). The physical mechanism responsible for the  asymptotic stability of Minkowski space is the dissipation by dispersion, that is the radiation of energy of perturbations to infinity.

 In the case of AdS, the question of stability must be supplemented by a choice of boundary conditions at infinity
 and,  a priori, an answer  may depend on the choice. Once this choice is made, one has to show that the initial-boundary value problem is locally well-posed; otherwise the question of stability does not make mathematical sense\footnote{I point out this obvious fact because sometimes, especially in numerical studies, the ill-posedness is mistaken with instability.}.
 Such a local well-posedness result for a large class of AdS boundary conditions was proved by Friedrich for vacuum Einstein's equations with negative cosmological constant in four dimensions \cite{f}. Here I will consider only
 so called reflective boundary conditions for which there is no flux of energy across the conformal boundary.
 In this case
the asymptotic stability of AdS is precluded because the conformal boundary acts like a mirror at which perturbations  propagating outwards bounce off and return to the bulk. This  leads to very complicated  nonlinear wave interactions  in the bulk,  understanding of which appears  challenging even for small perturbations. Thus, it is no wonder  that the question of stability of  AdS space remains open.

What is astonishing, however, is that until 2011 this basic question has been completely ignored in the avalanche of papers on asymptotically AdS spacetimes triggered by \cite{mal}.
 The only exception I know is the paper by M. Anderson \cite{a} in which he proved  that the only globally regular in time asymptotically AdS spacetime that tends to AdS for $t\rightarrow \pm \infty$ is the AdS space itself (this rigidity result is hardly surprising for a system that cannot lose energy to infinity). In the same paper Anderson cautiously conjectured stability of AdS by writing: ``One expects
that  $g_{AdS}$ is in fact dynamically stable, with the behavior of the nonlinear exact solutions
nearby to $g_{AdS}$ well-modeled on the linearized behavior." As far as I remember, this conjecture reflected the majority view\footnote{Not all seemed to share this view: in his talk at this workshop \cite{d}, M. Dafermos expressed some concerns whether stability is possible in the presence of non-decaying linearized perturbations.
However, these concerns were not based  on physical arguments for instability but rather on the lack of  conceivable mathematical methods of proving stability in the absence of dissipation by dispersion.
}
 at the workshop "Global Problems in Mathematical Relativity II" held at the Newton Institute (Cambridge) in October 2006, where the problem of stability of AdS was widely discussed and where I got interested in it.
\section*{Linear stability of $\text{AdS}$} Before addressing the hard question of nonlinear stability, let us consider a much simpler question of linear stability of AdS. For reflective boundary conditions this question reduces to a spectral problem for a certain master linear operator whose coefficients depend on the character (scalar, electromagnetic, or gravitational) of the perturbation. The pioneering studies of this problem by  Breitenlohner and Freedman  \cite{bf} (see also \cite{ais}) have been more recently extended and completed  by Ishibashi and Wald \cite{iw}. Let us summarize their results in the case of scalar perturbations, that is for the Klein-Gordon equation $\Box_g \psi-\mu^2 \psi=0$  on the AdS background.
After separation of angular variables
$\psi(t,x,\omega)=\sum\limits_{\vec{k}} \phi_{\vec{k}}(t,x)\, Y_{\vec{k}}(\omega)$ (where $Y_{\vec{k}}(\omega)$ are the scalar spherical harmonics on $S^{d-1}$), one gets
\begin{equation}\label{master}\partial_t^2 \phi + L \phi=0\,, \quad L=-\frac{1}{(\tan{x})^{d-1}}\, \partial_x \left((\tan{x})^{d-1} \, \partial_x\right)+\frac{\mu^2}{\cos^2{\!x}}+\frac{\ell(\ell+d-2)}{\sin^2{\!x}}\,,
\end{equation}
where $\ell$ is the degree of the spherical harmonic and the collective index $\vec{k}$ on $\phi_{\vec{k}}$ has been dropped.
One finds that for $\nu^2:= d^2/4+\mu^2>0$ solutions of this equation behave near $\mathcal{I}$ as follows (using $y=\pi/2-x$)
\begin{equation}\label{asym}
  \phi(t,x) \sim c_{+}(t)\, y^{\frac{d}{2}+\nu} + c_{-}(t)\, y^{\frac{d}{2}-\nu},
\end{equation}
which gives rise to three possible  boundary conditions at infinity, usually referred to as the Dirichlet ($c_-=0$), Neumann ($c_+=0$), or Robin ($c_++\gamma c_-=0$) boundary conditions.
 The operator $L$, defined on the Hilbert space $\mathcal{H}=L^2([0,\pi/2],(\tan{x})^{d-1} \,dx)$, is  positive for  $\nu^2\geq 0$. The lower limit $\nu^2=0$ corresponds to the well-known Breitenlohner-Freedman mass bound $\mu^2=-d^2/4$ \cite{bf} (in this case the second solution in \eqref{asym} falls off as $y^{d/2} \log{y}$). Below this bound (i.e. for $\nu^2<0$) there is no way to define unitary dynamics. For $\nu^2\geq 1$ the operator $L$  is essentially self-adjoint  and the Dirichlet boundary condition is forced by the requirement of square integrability.  For $\nu^2\leq 0<1$ there is a one-parameter freedom of choosing a self-adjoint extension, which amounts to choosing the Dirichlet, Neumann, or Robin boundary condition.

     This and an analogous result for electromagnetic and gravitational perturbations imply  that  the AdS space is linearly stable under scalar, electromagnetic, and gravitational perturbations
     obeying  the reflecting boundary conditions at infinity. The eigenvalue equation $L \phi=\omega^2 \phi$ is of hypergeometric type, hence the eigenmodes can be found explicitly. For example, in the case of Dirichlet boundary condition one finds
     the eigenvalues ($k=0,1,\dots$)
     \begin{equation}\label{eigen-val}
       \omega_k^2=(2k+\nu+\sigma+1)^2\,, \quad\mbox{where} \quad \sigma^2=\ell(\ell+d-2)+(d-2)^2/4\,.
     \end{equation}
     The
     corresponding orthonormal eigenfunctions are
     \begin{equation}\label{eigen-fun}
       e_k(x)= d_k \, (\cos{x})^{\frac{d}{2}+\nu} (\sin{x})^{\sigma+1-\frac{d}{2}}\,
       {}_{2}\text{F}_1(k+\nu+\sigma+1,-k,1+\sigma; \sin^2{\!x})\,,
     \end{equation}
     where $d_k$ is a normalization factor ensuring that $(e_j,e_k)=\delta_{jk}$ (hereafter, $(f,g)$ denotes the inner product in $\mathcal{H}$).
It follows from \eqref{eigen-val} that
    $\Delta\omega_k=2\Delta k$, hence the linearized waves  are nondispersive. This property will have important consequences for the nonlinear stability analysis.

Linearized dynamics \eqref{master} provides an accurate approximation of short time behavior of small perturbations of AdS and establishing linear stability is an important first step in understanding stability. However, the linear stability by no means implies (nonlinear) stability.
\section*{Nonlinear instability of $\text{AdS}_{d+1}$ for $d\geq 3$} The question of stability of AdS  in full generality seems to be out of reach of current PDE technology so it is natural to consider more tractable special cases, in particular spherically symmetric perturbations.
 Since by Birkhoff's theorem spherically symmetric vacuum solutions are static,
 one has to add matter to generate dynamics. A simple matter model is a self-gravitating minimally coupled massless scalar field whose dynamics is described by the Einstein-scalar field equations
\begin{equation}\label{einstein-scalar}
 G_{\alpha\beta} + \Lambda g_{\alpha \beta} =
 8 \pi G \left(\partial_{\alpha} \phi \,\partial_{\beta} \phi - \frac{1}{2}  g_{\alpha\beta} (\partial \phi)^2\right)\,,\qquad
 g^{\alpha\beta} \nabla_{\alpha} \nabla_{\beta} \phi=0\,.
\end{equation}
In the asymptotically flat case ($\Lambda=0$) the studies of this model have brought important insights, most notably the proof of  weak cosmic censorship by Christodoulou \cite{ch} and \nolinebreak the discovery of critical phenomena at the threshold for black hole formation by Choptuik \cite{matt}.

A few years ago, Andrzej Rostworowski and I have set out  to investigate the system \eqref{einstein-scalar} with $\Lambda<0$ in spherical symmetry.
We assumed the following parametrization of spherically symmetric asymptotically AdS spacetimes
\begin{equation}
g = \frac {1}{\cos^2{\!x}}\left( -A e^{-2 \delta} dt^2 + A^{-1} dx^2 + \sin^2{\!x} \, d\omega^2\right)\,,
\end{equation}
where $A$ and $\delta$ are functions of $(t,x)$. For this ansatz Eqs.\eqref{einstein-scalar} reduce to the quasilinear  system consisting of the scalar wave equation
\begin{equation}\label{wave}
\partial_t\left( A^{-1} e^{\delta} \partial_t \phi \right) = \frac{1}{(\tan{x})^{d-1}}\partial_x \left((\tan{x})^{d-1} \,A \, e^{-\delta} \partial_x \phi \right)\,,
\end{equation}
coupled to two ordinary differential equations equations (we set $8\pi G=d-1$)
\begin{equation}\label{cons}
\partial_x A = \frac{d-2+2\sin^2{\!x}} {\sin{x}\cos{x}} (1-A) - \sin{x}\cos{x}\, A\, \rho, \qquad
\partial_x \delta = - \sin{x}\cos{x}\, \rho\,,
\end{equation}
where $\rho=A^{-2} e^{2\delta} (\partial_t \phi)^2+(\partial_x \phi)^2$ is the scalar field energy density. Guided by the Schwarzschild-AdS solution \eqref{schw}, it is useful to define the mass function $m(t,x)$ by
$A(t,x)=1- m(t,x)(\cos{x})^d/(\sin{x})^{d-2}$.
 The requirement of no flux of mass through $\mathcal{I}$ enforces the Dirichlet asymptotics
 (using $y=\pi/2-x$)
\begin{equation}\label{bc}
    \phi(t,x)\sim y^d, \quad
    \delta(t,x)-\delta(t,\infty)\sim y^{2d},\quad
    A(t,x)-1 \sim y^d \,.
\end{equation}
For these boundary conditions, the total mass, defined as $M=\lim_{x\rightarrow \pi/2} m(t,x)$, is finite and conserved in time.
The system of equations \eqref{wave} and \eqref{cons} with the boundary conditions \eqref{bc} and compatible smooth initial data $(\phi,\partial_t\phi)_{|t=0}$ constitutes a
 locally well-posed  initial-boundary value problem  \cite{hs1}. Our investigations of global behavior of small data solutions to this problem have led us to the following conjecture (within the model, of course):
 \begin{conj}[\cite{br}] The $\text{AdS}_{d+1}$ space (for $d\geq 3$) is unstable  against the formation of a black hole for a large class of arbitrarily small perturbations.
    \end{conj}
In mathematical terms, a black hole is  detected by the formation of an apparent horizon at a radius $x_H$ where the metric function $A(t,x)$ drops to zero. Although our coordinate system breaks down at this point, it is clear from elsewhere (for instance, numerical simulations in horizon-penetrating coordinates or rigorous results \cite{hs1}) that eventually all the matter will fall inside the horizon and the spacetime will settle down to the Schwarzschild-AdS black hole \eqref{schw} with mass $M$ equal to the initial mass.

    The evidence for Conjecture~1, first given for $d=3$  \cite{br}  and later generalized to $d\geq 3$ \cite{jrb}, is based on perturbative and numerical calculations, which will be now summarized.
    \vskip 0.1cm
    \noindent\emph{Perturbative evidence.}
For small initial data
$(\phi,\partial_t\phi)_{|t=0}=(\ep f(x),\ep g(x))$, the dynamics of solutions   can be described (as long as the solutions remain small) using weakly nonlinear perturbation analysis.
 To this end  we expand the solution in the
perturbation series
\begin{equation}\label{expand}
\phi=\ep \phi_1+\ep^3 \phi_3+...,\quad
\delta=\ep^2 \delta_2+\ep^4 \delta_4+...,\quad
1-A=\ep^2 A_2+\ep^4 A_4+...
\end{equation}
where $(\phi_1,\partial_t\phi_1)_{|t=0}=(f(x),g(x))$ and $(\phi_j,\partial_t\phi_j)_{|t=0}=(0,0)$ for $j>1$.
 Inserting the expansion \eqref{expand} into the field equations \eqref{wave} and \eqref{cons} and collecting terms of the same order in $\ep$, we obtain a hierarchy of linear equations which can be solved order-by-order.
At the first order we get for $\phi_1$
the linear wave equation \eqref{master} with zero potential (because $\mu=0$ and $\ell=0$). In this case the eigenfrequencies \eqref{eigen-val} and the eigenfunctions \eqref{eigen-fun} simplify to
\begin{equation}\label{eigen}
  \omega_k=2k+d,\qquad e_k(x)= d_k \, (\cos{x})^{d}\,
       {}_2\text{F}_1(k+d,-k,d/2; \sin^2{\!x})\,.
\end{equation}
Thus, at the linear level the solution is
\begin{equation}\label{phi1}
\phi_1(t,x)=\sum_{k=0}^{\infty}  a_k \cos(\omega_k t+\beta_k) \, e_k(x),
\end{equation}
where the amplitudes $a_k$ and  phases $\beta_k$ are determined by the initial data. Using this solution at the second order we get from \eqref{cons} the perturbations of metric functions $A_2$ and $\delta_2$ (so called backreaction) and at the third order we obtain an inhomogeneous linear wave equation
 $\ddot \phi_3 + L \phi_3=S$, where the source $S$ depends on $\phi_1,A_2,\delta_2$ and their first derivatives.
 Projection of  this equation on the basis $\{e_k\}$ yields  an infinite system of decoupled forced harmonic oscillators for the generalized Fourier coefficients $c_k:=(\phi_3,e_k)$
 \begin{equation}\label{hierarchy}
   \ddot c_{k}+\omega_k^2\, c_{k} = S_k:=(S,e_k)\,.
 \end{equation}
 A calculation shows that each triple $(j,l,m)$ of indices of nonzero modes in the linearized solution \eqref{phi1} such that  $\omega_k=\omega_j+\omega_l-\omega_m$ gives rise to a resonant term in $S_k$ (i.e. a term proportional to $\cos{\omega_k t}$ or $\sin{\omega_k t}$). Note that this abundance of resonances is a consequence of the nondispersive character of the linearized spectrum. Some of the resonances may be removed by renormalizing the frequency, however the remaining resonances give rise to secular terms that grow linearly in time. A similar nonlinear perturbation analysis has been performed for the vacuum Einstein equations in \cite{dhs}.
  This breakdown of the perturbation expansion at the third order  signals the onset of instability at time of order $\ep^{-2}$.   We believe that the secular terms appearing in $\phi_3$ are progenitors of the higher-order resonant mode mixing which shifts the energy spectrum to higher frequencies.
 \vskip 0.1cm
 \noindent \emph{Numerical evidence.} The perturbative analysis is corroborated by numerical simulations which show that, indeed, generic perturbations start to grow rapidly after a time $\sim \ep^{-2}$ (see Fig.~2 in \cite{br}). This growth eventually leads to the formation of a horizon. On a heuristic level, the formation of the horizon is an expected consequence of the transfer of energy to high frequencies and, \emph{eo ipso}, small spatial scales.  Put differently, the formation of a black hole (in an amusing analogy to viscosity in fluid turbulence) provides a natural cutoff for the turbulent energy cascade. Our numerical results have been confirmed and extended to complex scalar fields by Buchel, Lehner, and Liebling \cite{bll1}.

 To demonstrate the transfer of energy to high frequencies we defined the Fourier coefficients $\Phi_k:=(A^{1/2}\,\partial_x \phi,\partial_x e_k)$ and $\Pi_k:=(A^{-1/2} e^{\delta}\,\partial_t \phi,e_k)$ and expressed the total  mass as  the Parseval sum $M=\sum_{k=0}^{\infty} E_k(t)$, where $E_k:=\Pi_k^2+\omega_k^{-2} \Phi_k^2$ is  the $k$-mode energy.
 The evolution of the energy spectrum, that is the distribution of mass among the modes,   is depicted in Fig.~2 for gaussian initial data.
  Initially, the energy is concentrated in low modes; the exponential
cutoff of the spectrum expresses the smoothness of initial
data. During the evolution the range of excited modes increases and
the spectrum becomes broader. Just before horizon formation the
spectrum exhibits the power-law scaling $E_{k}\sim k^{-\alpha}$, where the exponent $\alpha$  seems to be universal, i.e. the same for all collapsing solutions (but depending on dimension~$d$).  Clearly, the formation of the
power-law spectrum reflects  the loss of smoothness of the
solution during collapse, however we have not been able to compute $\alpha$~analytically.

 \begin{figure}[h]
    \includegraphics[width=0.9\linewidth]{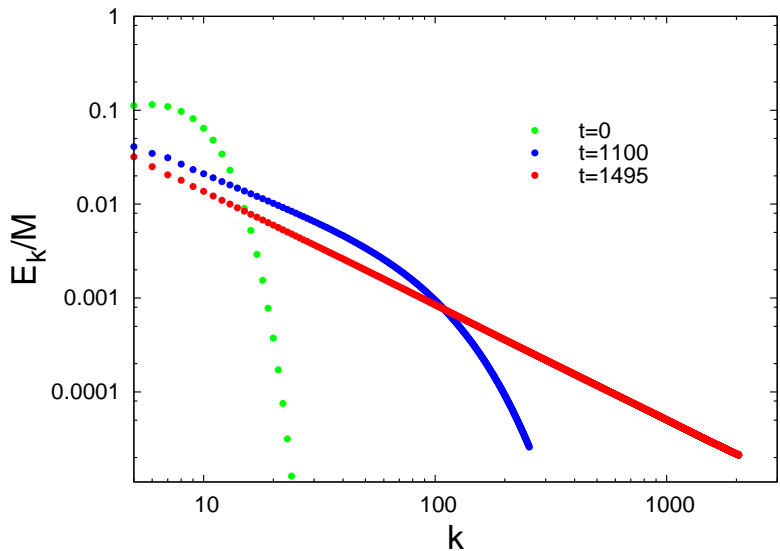}
    \caption{Log-log plot of the energy spectrum for $d=3$ at three moments of time: initial,
    intermediate, and just before collapse. The fit of power law
    $E_{k}\sim k^{-\alpha}$ at $t=1495$
    gives the slope~$\alpha\approx 1.2$.}
  \end{figure}

It should be stressed that the resonant transfer of energy is not active for some perturbations. For example, solutions starting from  one-mode initial data \cite{br} or one-mode-dominated initial data \cite{bll2} appear almost periodic  for a very long (possibly infinite) time. In particular, there is good  evidence for the existence of time-periodic solutions. The existence of such solutions (geons) was first conjectured by Dias, Horowitz and Santos \cite{dhs}  for the vacuum Einstein equations on the basis of perturbative calculations.

\begin{figure} [h]
    \includegraphics[width=0.4\linewidth]{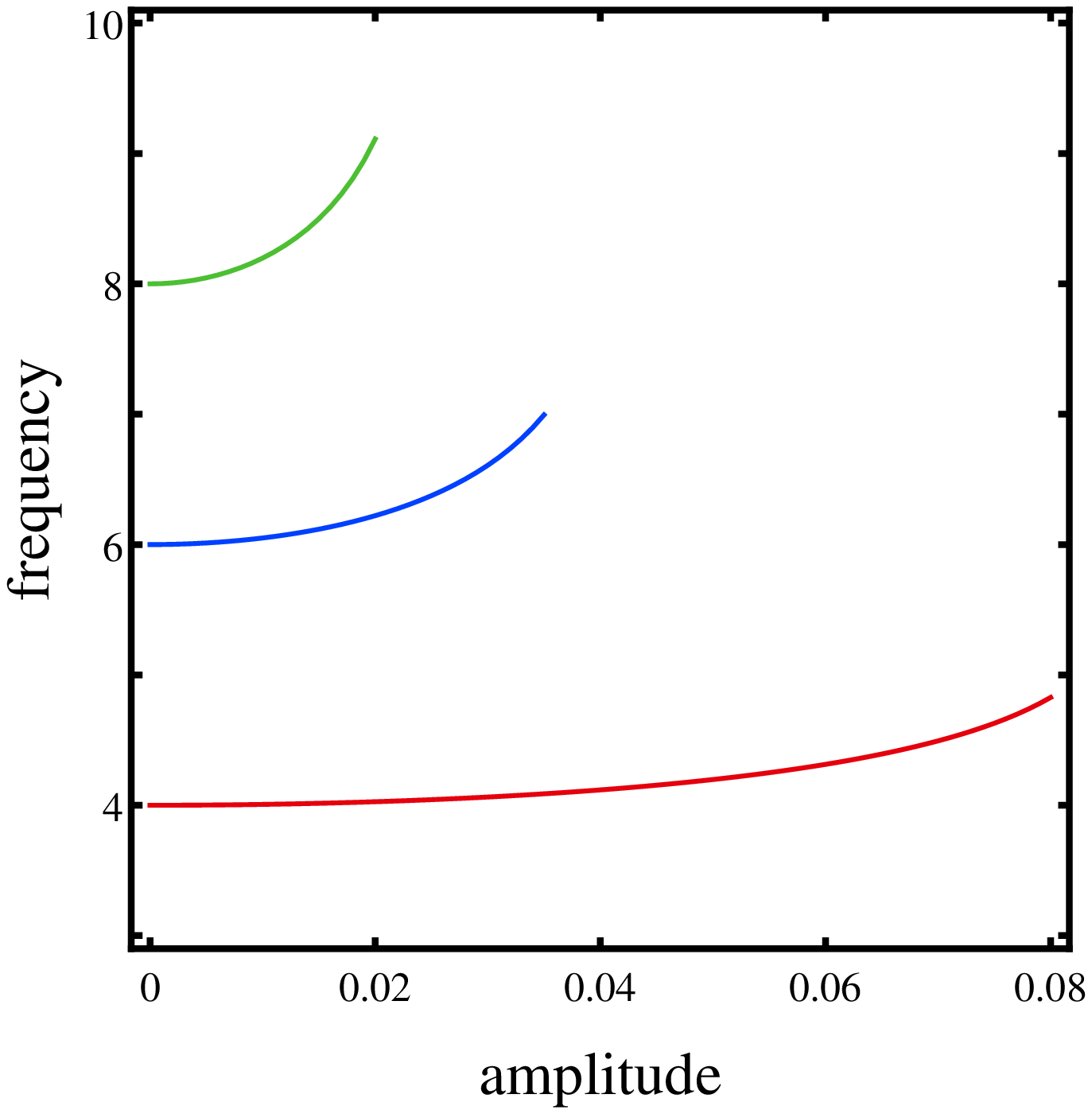}
    \includegraphics[width=0.48\linewidth]{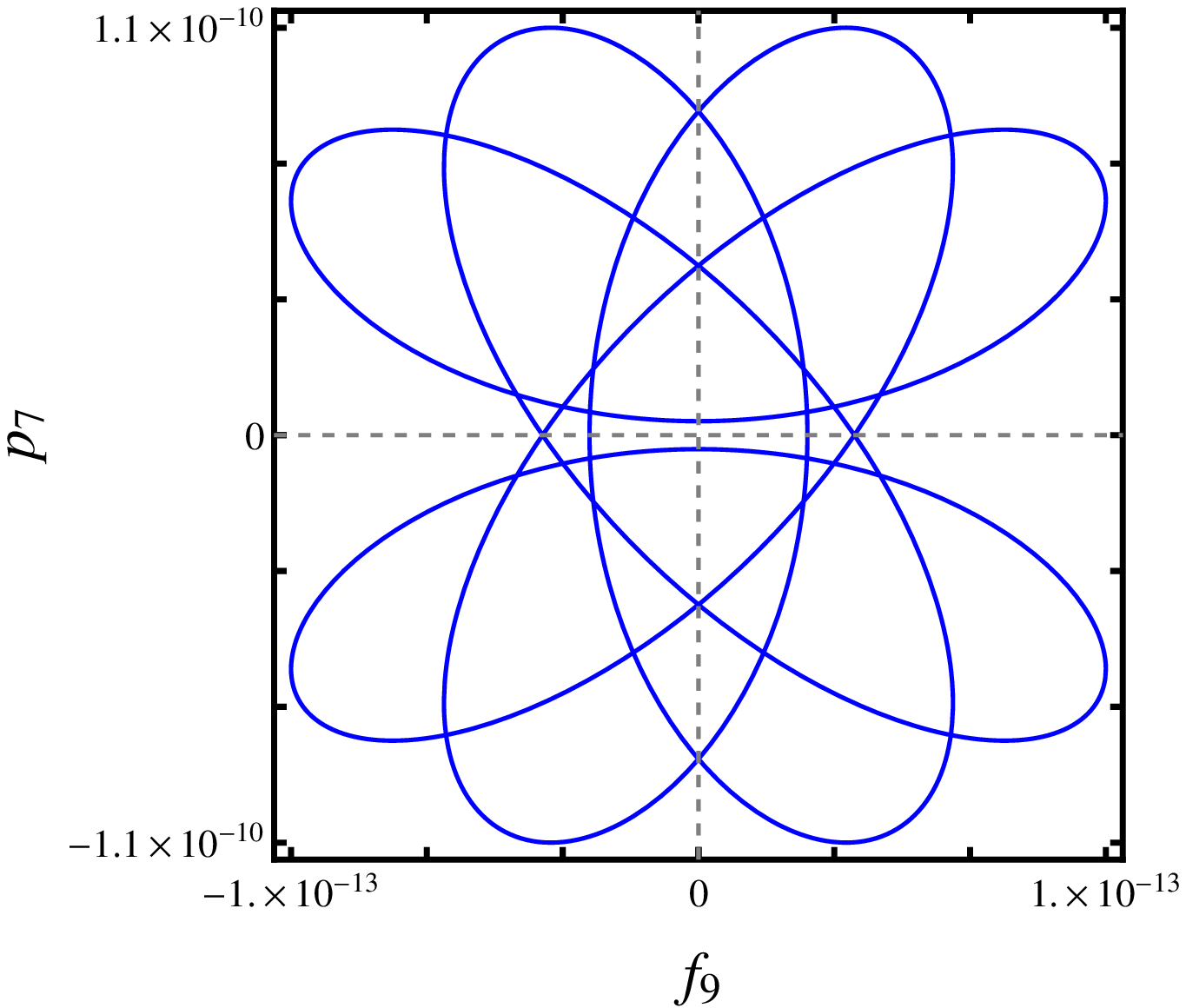}
    \caption{Left panel: the bifurcation diagram for the first three time-periodic solutions in $d=4$. Right panel: A two-dimensional projection of the phase trajectory is shown to remain closed after several hundreds of periods (see \cite{mr} for notation).}
  \end{figure}

For the Einstein-massless scalar model  time-periodic solutions, bifurcating from the eigenmodes on the linearized spectrum, were constructed by Maliborski and Rostworowski \cite{mr} in two independent ways: the Poincar\'e-Lindstedt perturbative method and a numerical spectral method. The outcomes of these two methods agree with great accuracy which leaves no reasonable doubt that time-periodic solutions are real entities. Moreover, numerical evolution of the corresponding initial data  retraces the periodic loop for hundreds of periods (see Fig.~3) which indicates that these time-periodic solutions are stable.

\section*{Nonlinear instability of $\text{AdS}_3$} Conjecture~1 is restricted to dimensions $d\geq 3$. The case of three-dimensional AdS gravity ($d=2$) is different for a very simple reason: for initial data with mass $M<1$ no apparent horizon can form because  $A(t,x)=1-m(t,x) \cos^2{\!x} >0$.  This is reflected in
 the mass gap\footnote{The existence of this finite mass threshold for black hole formation is a manifestation of the energy critical character of Einstein's equations in three dimensions.} between $\text{AdS}_3$ and the lightest black hole solution\footnote{AdS-Schwarzschild black holes in three dimensions are called BTZ black holes \cite{btz}.} \eqref{schw} with $M=1$. Since small perturbations of $\text{AdS}_3$  cannot evolve into black holes, we are left with the dichotomy: naked singularity formation  or global-in-time regularity. Recently, together with Joanna Ja\l mu\.zna  \cite{bj} we attempted to resolve this dichotomy for the Einstein-scalar field system \eqref{wave}-\eqref{cons} with the boundary conditions \eqref{bc}. We found that for typical initial data the dynamics is turbulent.
 The heuristic explanation of the mechanism which triggers the turbulent behavior is the same as in higher dimensions, namely the generation of secular terms by resonant four-wave  interactions.
  Actually, in three dimensions the rate of transfer of energy to high frequencies is much faster than in higher dimensions which  puts stringent demands on the spatial resolution and severely limits the times accessible in  numerical simulations.

   In order to extract  information about regularity of solutions  from numerical data, we used the analyticity strip method due to Sulem, Sulem, and Frisch \cite{ssf}. This method is based on the following idea. Consider a solution $u(t,x)$ of some nonlinear evolution equation for real-analytic initial data and let $u(t,z)$ be its analytic extension to the complex plane of the spatial variable. Typically, $u(t,z)$ will have complex singularities (coming in complex-conjugate pairs) which move in time.
The imaginary part of the complex singularity $z=x+i \rho$ closest to the real axis determines the radius of analyticity of the solution. Monitoring the time evolution of $\rho(t)$ and checking if it vanishes (or not) in finite time, one can predict (or exclude) the blowup. The key observation is that the value of $\rho$ is encoded in  the asymptotic behaviour of  Fourier coefficients of $u(t,x)$ which decay exponentially as $\exp(-\rho k)$ for large~$k$ (with an algebraic prefactor depending on the type of the singularity). Therefore, the analyticity radius $\rho(t)$ can be obtained by fitting an exponential decay to the tail of the numerically computed Fourier spectrum. Applying  this technique to the problem at hand we have arrived at the following conclusion:
\begin{conj}[\cite{bj}]
 Small smooth perturbations of $\text{AdS}_3$ remain smooth for all times  but their radius of analyticity shrinks to zero exponentially fast.
\end{conj}
The evidence supporting this conjecture is summarized in Figs.~4 and 5 generated from the numerical evolution of small gaussian perturbations. In Fig.~4, showing the evolution of  energy spectrum, one can see that the range of frequencies participating in the evolution increases very rapidly but, in contrast to Fig.~2, no power-law behavior is seen.
In accord with the analyticity strip method we assumed that for large wave numbers the energy spectrum is described by the formula
$
   E_k(t)=C(t)\,k^{-\beta(t)} e^{-2 \rho(t) k}
$.
 Fitting this formula to the numerical data  we found that after some transient period of  time the radius of analyticity $\rho(t)$ is well approximated by the exponential decay
 $\rho(t)=\rho_0 \, e^{-t/T}$ with
  the characteristic decay time $T$ scaling as $\varepsilon^{-2}$, where $\ep$ is the amplitude of perturbation. The good quality of this fit (see Fig.~3 in \cite{bj}) suggests (with a little dose of optimism) that the exponential decay can be extrapolated forever, thereby justifying Conjecture~2.

      \begin{figure}[h]
 \includegraphics[width=0.47\textwidth,angle=270]{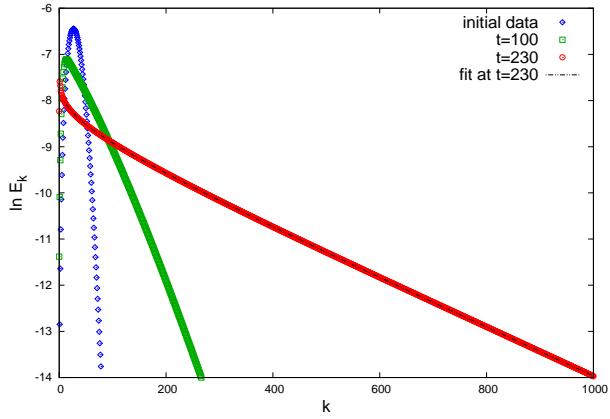}
  \caption{Time evolution  of the energy spectrum for a small gaussian perturbation.}
 \label{log_dec}
\end{figure}

    The exponentially fast shrinking of the radius of analyticity is reflected in the exponentially fast growth of  Sobolev norms $\dot H_s$ of the scalar field for $s>1$, as illustrated in Fig.~5 for $s=2$. After an initial quiescent period, whose duration scales as $\varepsilon^{-2}$, the maxima of $\dot H_2(t)$ begin to grow exponentially  approximately as  $\exp(t/T)$. Thus, even though
  smooth  initially  small perturbations  remain smooth forever, they
do not remain small in any reasonable norm that captures the turbulent behavior, which means that $\text{AdS}_3$ is unstable.
This kind of gradual loss of regularity, where  solutions develop progressively finer spatial scales as $t\rightarrow \infty$ without ever losing smoothness (sometimes referred to as weak turbulence), has been  well known in fluid dynamics \cite{y}.

 \begin{figure} [h]
 \includegraphics[width=0.47\textwidth,angle=270]{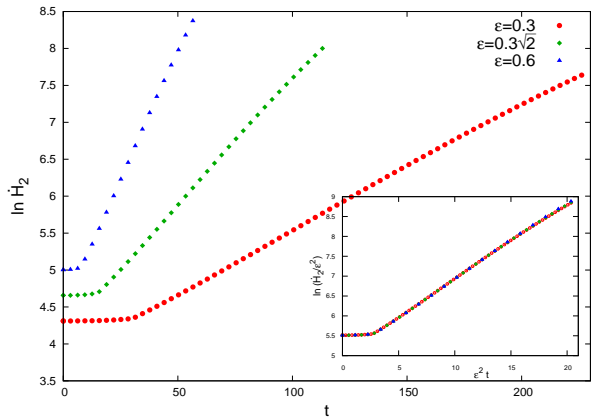}
  \caption{Time evolution of the Sobolev norm $\dot H_2=\|\partial^2_{x}\phi(t,x)\|_2 $ for gaussian perturbations with three different small amplitudes $\ep$ (only the upper envelope of rapid oscillations is plotted). In the inset,  the curves  are shown to coincide after rescaling $t\rightarrow \ep^2 t, \dot H_2 \rightarrow \ep^{-2} \dot H_2$.}
 \label{norm}
\end{figure}
\section*{Concluding remarks}
The attempts to answer the question raised in the title have opened up new and unexpected research paths lying at the interface of general relativity and theory of turbulence, which is pretty much an uncharted territory. Admittedly, the results presented above raise more questions than giving answers.  Let me conclude by mentioning some of the most interesting open questions (besides, of course, proving or refuting Conjectures~1 and 2):
\begin{itemize}

\item  What is the role of negative cosmological constant $\Lambda$ in the observed phenomena? Is an extra attractive force due to $\Lambda$ important in triggering instability of AdS, or
  is the only role of $\Lambda$ to confine the evolution in an effectively bounded domain? Some evidence for the latter  was given by Maliborski who observed a similar turbulent  instability for flat space inside a spherical cavity with perfectly reflecting walls \cite{m}.

\item Is the nondispersive character of the linearized spectrum essential for the turbulent instability? In other words, how important are exact resonances as a driving mechanism of the turbulent cascade (see \cite{dhms} for an interesting discussion of this issue). In attempting to answer this very difficult question, touching upon the celebrated KAM theorem and the problem of small denominators, one should keep in mind that an answer is not known  for such simple equations as  the cubic wave equation $\partial^2_{t} \phi-\partial^2_{x} \phi +\mu^2 \phi+\phi^3=0$ on the interval $0\leq x\leq \pi$ with Dirichlet boundary conditions. Note that in this case the linear dispersion relation is $\omega_k=\sqrt{k^2+\mu^2}$, hence the linearized waves are nondispersive only for $\mu=0$ (in contrast to the linearized waves on AdS  which are nondispersive for any $\mu$, as follows from \eqref{eigen-val}).

\item What determines the exponent in  the power-law energy spectrum of the turbulent cascade in $d\geq 3$?

\item What determines a borderline of stability islands around time-periodic solutions? This is closely related to the question of how generic is the turbulent instability.

\item What happens outside spherical symmetry? In particular, what is the endpoint of instability of $\text{AdS}_{d+1}$ in $d\geq 3$ for non-spherical perturbations? The answer is far from obvious, because it is not clear if a natural candidate for the endstate, the Kerr-AdS black hole, is stable itself (possible obstructions to stability  being due to superradiance and stable trapping phenomena \cite{hs2}).

 \item What is the nature of the threshold for black hole formation in $d=2$? Numerical investigations of this  question by Pretorius and Choptuik \cite{pc} provided important insights into the near-critical dynamics, however the critical solution itself remains not understood \cite{carsten}. Does every solution with $M>1$ evolve into a black hole?
\end{itemize}

In conclusion, I cannot resist noting that the results described above demonstrate once again that from numerical explorations of Einstein's equations one can gain understanding of phenomena which would hardly be possible by purely analytic means. The computer, as an astronomer's telescope,  allows us to see things that otherwise could have remained hidden.
The role of computation in general relativity seems destined to expand in future.

\begin{acknowledgements}
It is  a pleasure to thank Andrzej Rostworowski, Tadeusz Chmaj, Joanna Ja\l mu\.zna, and Maciej Maliborski for collaboration on various aspects of the work described here. I am grateful to Helmut Friedrich for teaching me many things about AdS. I also thank Lars Andersson, Piotr Chru\'sciel, Mihalis Dafermos, Gary Gibbons, Carsten Gundlach, Pawe\l{} Mazur, and Robert Wald for helpful remarks. This work was supported by the NCN grant DEC-2012/06/A/ST2/00397.
\end{acknowledgements}

\end{document}